# Engineering soliton nonlinearities: From local to strongly nonlocal

Yaroslav V. Kartashov,[1] Victor A. Vysloukh,[2] and Lluis Torner[1]

[1]ICFO-Institut de Ciencies Fotoniques, and Universitat Politecnica de Catalunya,
Mediterranean Technology Park, 08860 Castelldefels (Barcelona), Spain

[2]Departamento de Fisica y Matematicas, Universidad de las Americas – Puebla, Santa Catarina Martir, 72820, Puebla, Mexico

We put forward a strategy to achieve synthetic nonlinearities where local and nonlocal contributions compete on similar footing, thus yielding intermediate tunable responses ranging from fully local to strongly nonlocal. The physical setting addressed is a semiconductor material with both Kerr and thermal nonlinearities illuminated by a pulse train with suitable single-pulse width and repetition rate. We illustrate the potential of the scheme by showing that it supports soliton properties that are not accessible in either limit of purely local or purely nonlocal media.

OCIS codes: 190.0190, 190.6135

The global nature of the nonlinear response of a given material is a central property that determines the potential phenomena exhibited by an intense light beam propagating in it. A salient example is provided by nonlocal versus local nonlinearities, a difference that may lead to drastically different light evolution. For example, in contrast to uniform materials with purely local cubic nonlinearities where multidimensional and multipole solitons are highly prone to instabilities and vortex solitons are azimuthally unstable, nonlocal nonlinearities arrest collapse [1], the character of soliton interactions changes dramatically [2,3] making possible the existence of bound soliton states [4-8], and vortex solitons can be stabilized [9-11]. The possibility to tune nonlinearity from predominantly local to predominantly nonlocal is possible is suitable materials, including liquid crystals [12], while competition of nonlinearities with different nonlocality degrees was considered theoretically in phenomenological models [13,14]. This is an important and potentially far-reaching program, because access to intermediate nonlinear responses should open a new toolkit to control the propagation of nonlinear waves rendering phenomena not possible at the limiting cases.

In this Letter we put forward a new strategy to explore such intermediate local-nonlocal response. We consider a semiconductor material, such us AlGaAs [15-19], where the contribution to the refractive index $\delta n \sim 10^{-4}$ due to the Kerr effect requires peak intensity of the order of $I \sim \text{GW/cm}^2$. Note that with thermal nonlinearities a similar refractive index change might be achieved by heating with a CW beam carrying $I \sim 100 \text{ W/cm}^2$ focused down to a spot size of $\sim 10$ $\mu$m. Even though soliton experiments are typically conducted with pulsed light [18,19], the thermal contribution can be neglected because a single short pulse carries a too small energy to heat the sample in a significant way. By tuning the pulse duration and repetition rate, here we study scenarios where the strength of local Kerr and nonlocal thermal nonlinearities become comparable thus yielding an intermediate local-nonlocal nonlinearity. We show, in particular, that such synthetic response supports multipole solitons that are stable for any number of poles, a phenomenon that does not occur in either of the local or nonlocal limiting cases.

We address the propagation of a light beam along the $\xi$ axis of an optical waveguide of width $L$, whose left and right edges located at $\eta = \pm L/2$ are thermo-stabilized by means of an external heat sink. The beam diffracts along the $\eta$ axis and is slightly absorbed upon propagation, thus acting as a heat source that generates an inhomogeneous temperature distribution. Thus, the light beam experiences both, Kerr and thermal nonlinearities. We assume pulse-light illumination, a key ingredient to control the relative strength of the Kerr and thermal nonlinearities. Under these conditions light propagation is described by the equation for the field amplitude $q$ coupled to the thermal conductivity equation for the temperature distribution $\theta$ given by:

$$\begin{aligned} i\frac{\partial q}{\partial \xi} &= -\frac{1}{2}\frac{\partial^2 q}{\partial \eta^2} - |q|^2 q - \gamma \theta q, \\ \frac{\partial \theta}{\partial \tau} - \frac{\partial^2 \theta}{\partial \eta^2} &= \phi(\tau)|q|^2. \end{aligned} \tag{1}$$

Here $\xi = z/k_0 x_0^2$ is the propagation distance normalized to the diffraction length $k_0 x_0^2$; $\eta = x/x_0$ is the transverse coordinate normalized to the transverse scale $x_0$; $q = (k_0 x_0^2 \omega n_2 / c)^{1/2} E$ is the field amplitude; $n_2$ is the Kerr nonlinearity coefficient; $\tau = t\kappa/x_0^2$ is the normalized time; $\kappa$ is the thermo-diffusion coefficient; $\theta = (\rho C_\text{p} \kappa \omega n_2 k_0 / \alpha c)T$ is the normalized temperature; $\rho$ is the density; $C_\text{p}$ is the specific heat capacity; $\alpha$ is the absorption coefficient; $\gamma = \alpha \beta x_0^2 / n_2 \rho C_\text{p} \kappa$ is a coefficient that stands for the relative strength of the thermal nonlinearity; and $\beta$ is the thermo-optic coefficient. The equation for $\theta$ is solved with boundary conditions $\theta|_{\eta=\pm L/2} = 0$. We assume that the function $\phi(\tau)$ describes a train of Gaussian pulses $\exp(-\tau^2/\tau_{\rm dur}^2)$ with width $\tau_{\rm dur}$ and repetition rate $\tau_{\rm rep} \gg \tau_{\rm dur}$. For AlGaAs under typical conditions one has $\alpha \approx 0.2$ cm$^{-1}$, $n_2 \approx 1.5 \times 10^{-13}$ cm$^2$/W at $\lambda = 1.55\,\mu$m, $\beta \approx 2.7 \times 10^{-4}$ K$^{-1}$, $\rho = 5$ g/cm$^3$, $C_{\rm p} \approx 0.33$ J/gK, $\kappa \approx 0.1$ cm$^2$/s. Thus, $x_0 = 10\,\mu$m yields $x_0^2/\kappa \approx 10^{-5}$ s, $\alpha c/\rho C_{\rm p}\kappa \omega n_2 k_0 \approx 1440$ K, and $\gamma = 21.6$. For these parameters, an on-axis increase of temperature of the order of 30 K corresponds to $\gamma\theta \approx 0.45$. We set $L = 20$.

The idea behind our scheme is illustrated in Fig. 1. It shows the spatio-temporal temperate distribution in a sample heated with pulse train. When each pulse arrives it raises the temperature in the region of highest intensity. The peak intensity of each pulse is assumed to be sufficiently high, so that around the pulse peak the Kerr nonlinearity is significant. While the Kerr nonlinearity is determined only by the peak intensity, the temperature distribution is dictated by the pulse energy. When $\tau_{\rm dur}$ is small the temperature increase generated by each pulse is small. After each pulse the temperature decays, so that for sufficiently large $\tau_{\rm rep}$ the temperature $\theta$ between subsequent pulses decreases to a negligible values. However, as $\tau_{\rm dur}$ increases and $\tau_{\rm rep}$ decreases, a significant thermal lens builds up inside the sample, since temperature $\theta$ does not decrease between pulses. Such thermal lens impacts beam evolution with a strength that may be comparable to that caused by the Kerr nonlinearity. Thus, by properly varying the parameters of the pulse train one may engineer the overall response of the material from fully local to strongly nonlocal. Here we consider the parameter range where the two nonlinearities compete on similar footing.

To get insight into the effects caused by the competing local-nonlocal nonlinearities, at first approximation one may use in the first of Eqs. (1) the temperature taken at the pulse peak $\tau = 0$. This is justified as long as the maximum on-axis temperature variation $\theta_{\rm d} = (\theta_{\max} - \theta_{\min})|_{\eta=0}$ is not too large in comparison to $\theta_0 = \theta|_{\eta=\tau=0}$. This allows obtaining stationary soliton solutions of Eq. (1) that can be written in the form $q = w(\eta)\exp(ib\xi)$, with $b$ being the propagation constant. The temporal intensity distribution is always given by $\phi(\tau)I = \phi(\tau)w^2$. Figure 2 shows typical soliton shapes (note that the beam is only trapped in space) together with corresponding temperature distributions.

The presence of nonlocal thermal nonlinearities affords the existence of multipole solitons, which can not exist in purely Kerr materials because such states are composed of several out-of-phase bright spots that repel each other. The simplest dipole soliton is shown in Figs. 2(c) and 2(d), but a variety of more complicated states of this type with larger num-ber of poles can be found. One can see that with increasing peak intensity, the on-axis temperature $\theta_0$ and its variation $\theta_d$ grows [Figs. 2(a) and 2(c)]. The temperature grows in a noticeable way only in the region where the pulse intensity is significant and slowly decreases between subsequent pulses. It should be stressed that the spatial temperature distributions differ considerably from those encountered in steady-state in thermal media, where $\theta$ always increases monotonically toward the beam center, even for higher-order multipoles. In contrast, in Figs. 2(b) and 2(d) one can see appearance of local maxima (whose positions overlap with intensity maxima) on top of the otherwise smooth and broad temperature distributions. Note that close to the sample boundaries at $\eta = \pm L/2$ the temperature decreases almost linearly. This is a direct consequence of the transient nature of thermal response generated by the pulse train. The local maxima become more pronounced as the temporal temperature variation $\theta_d$ increases, i.e. as $b$ grows. In the case of multipoles the separation between intensity maxima decreases with increasing peak intensity.

The properties of the fundamental solitons of Eq. (1) obtained using the model described above are summarized in Fig. 3. The energy flow, defined as $U = \int_{-\infty}^{\infty} |q|^2 \, d\eta$, increases monotonically with $b$. At fixed $b$ the energy flow attains its maximum value in the limit $\tau_{\text{dur}} \to 0$ when the thermal nonlinearity vanishes and $w \to (2b)^{1/2} \text{sech}[(2b)^{1/2}\eta]$. With increasing $\tau_{\text{dur}}$ the corresponding additional focusing thermal contribution becomes more and more pronounced, and causes a decrease of the peak intensity (hence $U$) necessary for stationary propagation [Fig. 3(a)]. Similarly, decreasing the repetition rate $\tau_{\text{rep}}$ also results in a decrease of the soliton energy flow at a fixed $b$. We found that at $b = 1$, with a fixed $\tau_{\text{dur}}$ the on-axis temperature $\theta_0$ corresponding to stationary solution of Eq. (1) is a monotonically decreasing function of $\tau_{\text{rep}}$, while the on-axis temperature variation $\theta_d$ first increases and then asymptotically approaches a constant value as $\tau_{\text{rep}} \to \infty$ [Fig. 3(b)]. As expected, when $\tau_{\text{rep}}$ decreases the temperate does not decrease significantly between subsequent pulses, an effect that results in the increase of the stationary part of the temperature distribution as well as in an increase of $\theta_0$. This makes the thermal nonlinearity stronger so that the soliton amplitude with fixed $b$ decreases and each pulse carries a smaller energy, resulting in a lower temperature variation $\theta_d$ upon passing of each single pulse. Hence, in this regime the thermal response dominates over the Kerr effect and it is reminiscent to the thermal response of the usual steady-state case.

In the other limit, when $\tau_{\text{rep}} \to \infty$, the temperature relaxes almost completely between subsequent pulses, the stationary part of the temperature diminishes and $\theta_d$ is determined by the energy carried by each pulse. In this case the soliton amplitude and $\theta_d$ asymptoti-cally approach constant values. Notice that when one keeps fixed $U$ instead of $b$, the temperature variation $\theta_\mathrm{d}$ does not change with $\tau_\mathrm{rep}$, while $\theta_0$ monotonically decreases with $\tau_\mathrm{rep}$. Increasing the pulse duration at a fixed $b$ and $\tau_\mathrm{rep}$ results in a monotonic growth of the on-axis temperature and the temperature variation [Fig. 3(c)]. The limit $\tau_\mathrm{dur} \to 0$ corresponds to negligible heating of the material and domination of local Kerr nonlinearity. These results clearly show the possibility to control the relative weights of Kerr and thermal nonlinearities by varying $\tau_\mathrm{dur}$ and $\tau_\mathrm{rep}$.

An example of the new phenomena made possible by the competing local-nonlocal nonlinearities is in order. On physical grounds, the specific spatial temperature distribution featuring local maxima around the soliton peaks may affect profoundly the soliton stability. While stability of fundamental solitons is this system is not surprising, we found that multipoles can be stable too, *for unlimited number of poles*. This is in clear contrast to both limits of purely local Kerr nonlinearity, where multipoles do not even exist, and of purely nonlocal steady-state thermal nonlinearity, where the number of stable poles was found to be limited to a maximal number [5]. Examples of stable propagation of two-hump and five-hump solitons (perturbed by the addition of initial random spatial noise) are shown in Figs. 4(a) and 4(b) for parameter values where the Kerr and the thermal nonlinear contributions amount to comparable values. We tested multipoles containing up to 10 poles, and found that all of them appear to be stable.

We thus conclude by stressing that the scheme put forward here affords the possibility to realize intermediate nonlinear responses where local and nonlocal components compete on similar footing. Such complex response may give rise to new physical phenomena not accessible in purely local or purely nonlocal media.

# References with titles

# References without titles

# Figure captions

Figure 1. Typical spatiotemporal temperature distributions upon heating of sample with pulse sequence when transverse intensity distribution features one (a) or two (b) humps. In both cases $\tau_{\rm rep} = 50$, $\tau_{\rm dur} = 0.5$.

Figure 2. Temporal distributions (a),(c) and spatial distributions at $\tau = 0$ (b),(d) of intensity $\phi I$ (black curves) and temperature $\gamma\theta$ (red curves) for fundamental (top row) and dipole (bottom row) solitons at $\tau_{\rm rep} = 5$ and $\tau_{\rm dur} = 0.004$. Upper curves in each panel correspond to $b = 2$, while lower curves correspond to $b = 1$. For fundamental solitons temporal distributions are shown at $\eta = 0$, while for dipoles they are shown in the point corresponding to intensity maximum.

Figure 3. (a) Energy flow versus propagation constant for fundamental soliton at $\tau_{\rm rep} = 5$ for $\tau_{\rm dur} \to 0$ (1), $\tau_{\rm dur} = 0.004$ (2), and $\tau_{\rm dur} = 0.009$ (3). Points marked by circles correspond to solitons in Figs. 2(a) and 2(b). Temperature $\gamma\theta_0$ at $\eta = \tau = 0$ and maximal temperature difference $\gamma\theta_{\rm d}$ at $\eta = 0$ versus period of pulse sequence at $\tau_{\rm dur} = 0.002$ (b) and versus pulse duration at $\tau_{\rm rep} = 5$ (c). In (b) and (c) $b = 1$.

Figure 4. Stable propagation of perturbed dipole (a) and five-hump (b) solitons with $b = 4$, $\tau_{\rm rep} = 5$, $\tau_{\rm dur} = 0.004$.

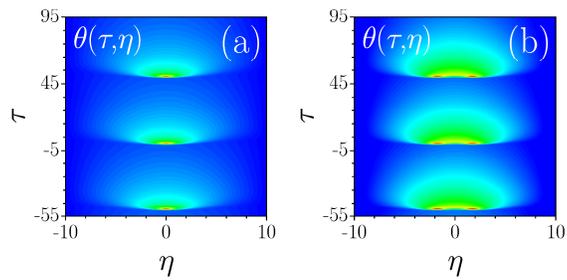

Figure 1.   Typical spatiotemporal temperature distributions upon heating of sample with pulse sequence when transverse intensity distribution features one (a) or two (b) humps. In both cases $\tau_{\rm rep}=50$, $\tau_{\rm dur}=0.5$.

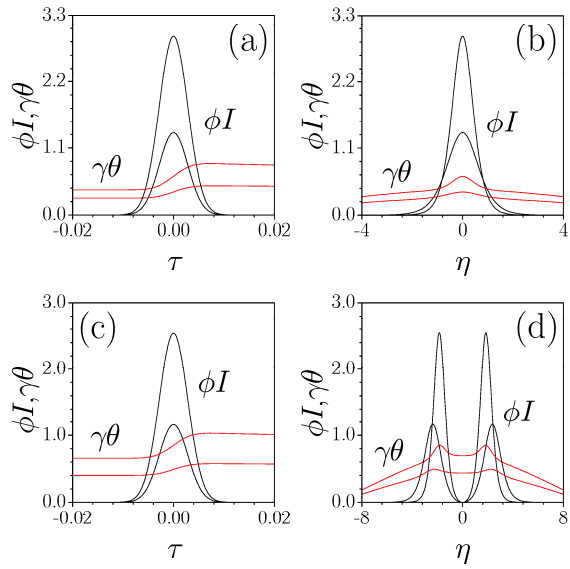

Figure 2. Temporal distributions (a),(c) and spatial distributions at $\tau = 0$ (b),(d) of intensity $\phi I$ (black curves) and temperature $\gamma\theta$ (red curves) for fundamental (top row) and dipole (bottom row) solitons at $\tau_{\rm rep} = 5$ and $\tau_{\rm dur} = 0.004$. Upper curves in each panel correspond to $b = 2$, while lower curves correspond to $b = 1$. For fundamental solitons temporal distributions are shown at $\eta = 0$, while for dipoles they are shown in the point corresponding to intensity maximum.

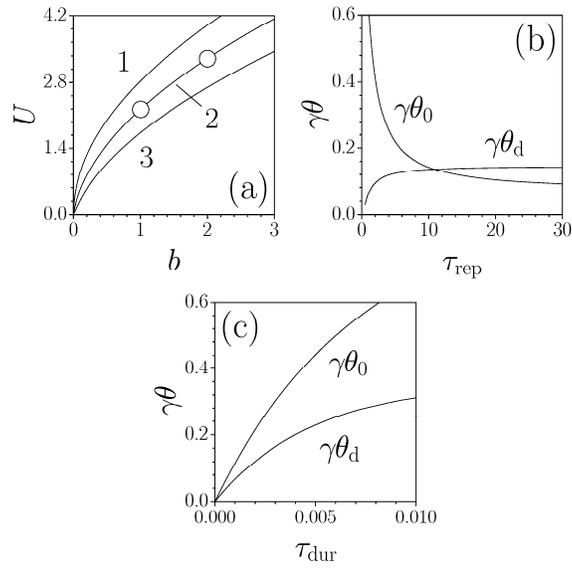

Figure 3. (a) Energy flow versus propagation constant for fundamental soliton at $\tau_{\text{rep}} = 5$ for $\tau_{\text{dur}} \to 0$ (1), $\tau_{\text{dur}} = 0.004$ (2), and $\tau_{\text{dur}} = 0.009$ (3). Points marked by circles correspond to solitons in Figs. 2(a) and 2(b). Temperature $\gamma\theta_0$ at $\eta = \tau = 0$ and maximal temperature difference $\gamma\theta_{\text{d}}$ at $\eta = 0$ versus period of pulse sequence at $\tau_{\text{dur}} = 0.002$ (b) and versus pulse duration at $\tau_{\text{rep}} = 5$ (c). In (b) and (c) $b = 1$.

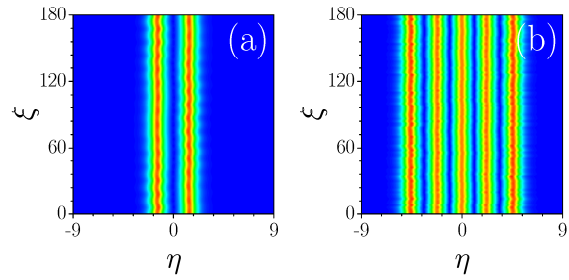

Figure 4. Stable propagation of perturbed dipole (a) and five-hump (b) solitons with $b = 4$, $\tau_{\text{rep}} = 5$, $\tau_{\text{dur}} = 0.004$.